\documentclass[prl,showpacs,twocolumn]{revtex4}

\usepackage{graphicx}
\usepackage{dcolumn}
\usepackage{bm}

\begin{document}

\preprint{APS/123-QED}

\title{First direct measurement of optical phonons in 2D plasma crystals}

\author{L. Cou\"edel}
\email{lcouedel@mpe.mpg.de}
\author{V. Nosenko}%
\author{S. K. Zhdanov}
\author{A. V. Ivlev}
\author{H. M. Thomas}
\author{G. E. Morfill}
\affiliation{%
Max-Planck-Institut f\"ur extraterrestrische Physik, 85741 Garching, Germany
}%

\date{\today}

\begin{abstract}
Spectra of phonons with out-of-plane polarization were studied experimentally in a 2D plasma crystal. The dispersion
relation was directly measured for the first time using a novel method of particle imaging. The out-of-plane mode was proven
to have negative optical dispersion, comparison with theory showed good agreement. The effect of the plasma wakes on the
dispersion relation is briefly discussed.
\end{abstract}


\maketitle

Complex plasmas consist of particles immersed in a weakly ionized gas-discharge plasmas \cite{Fortov05,1063-7869-47-5-R02,couedelepl2008}. Due to the
absorption of ambient electrons and ions, microparticles acquire significant (negative) charges and can form
strongly coupled systems analogous to conventional soft matter (complex fluids). Two-dimensional (2D) complex plasmas are
particularly convenient systems for the detailed experimental studies of strongly coupled phenomena at the atomistic level.
The essential advantage of such systems is that one can gain complete information about the state of the entire particle
ensemble in the phase space, which is an invaluable advantage for the investigation of collective processes occurring in
strongly coupled media, such as transport, phase transitions, structural quenching, etc \cite{Fortov05,1063-7869-47-5-R02}.

2D complex plasmas are usually studied in ground-based experiments with RF discharges. Microparticles injected in such
discharges levitate in the sheath region near the bottom electrode, where the electric field can balance  gravity. Under
certain conditions the particles can form a monolayer, and if the electrostatic coupling between them is strong enough,
particles arrange themselves into ordered structures -- 2D plasma crystals \cite{PhysRevLett.73.652}. In such crystals,
numerous collective phenomena like melting and recrystallization, mass and heat diffusion, solitons and shocks
\cite{popel:056402,PhysRevLett.89.035001,PhysRevLett.84.5141,nunomura:nzsm,PhysRevLett.86.2569,nosenko:nzikm}, etc. have
been reported. Propagation of  dust-lattice (DL) waves in 2D lattices is often used as a diagnostic tool to determine
parameters of the plasma crystal \cite{PhysRevLett.86.2569,PhysRevLett.88.215002}.

In 2D complex plasmas, like in any (strongly coupled) 2D system, two \textit{in-plane} wave modes can be sustained (here we
naturally leave aside polydisperse mixtures). In crystals, both modes have an acoustic dispersion, one of them is
longitudinal, another is transverse. Since the strength of the vertical confinement in such systems is finite, there is a
third fundamental wave mode associated with the \textit{out-of-plane} oscillations \cite{PhysRevE.63.016409,samsonov:026410,
PhysRevE.56.R74,vladimirov:030703,PhysRevLett.71.2753,PhysRevE.68.046403,zhdanov:zim}. Theory predicts that this mode has a negative (or inverse) optical
dispersion and depends critically on the parameters of the plasma wakes (that are formed downstream from each particle
levitating in the sheath \cite{PhysRevE.63.016409,zhdanov:zim}). So far, the systematic experimental studies of the DL waves
were limited by the in-plane modes, whereas direct measurements of the \textit{out-of-plane} dispersion relation have never
been carried out (see, however, Refs. \cite{PhysRevE.68.026405,samsonov:026410}). To a large extent, this is due to the
lack of reliable diagnostics of the vertical dynamics of individual particles.

In this Letter, we report on the first direct measurements of the \textit{out-of-plane} DL wave mode in a 2D plasma crystal.
The dispersion relation is recovered in a broad range of wave vectors, revealing remarkably good agreement with the existing
theory. The measurements were performed by employing a new detection technique of the vertical (out-of-plane) particle
displacement. The obtained results provide us with new insights into the mechanisms governing the individual particle
dynamics in 2D complex plasmas (in particular, highlight the important role of the plasma wakes) and allow us to specify the
conditions when such systems can be used to investigate generic properties of (classical) 2D solids and liquids.

\begin{figure}
\centering
\includegraphics[width=7.8cm]{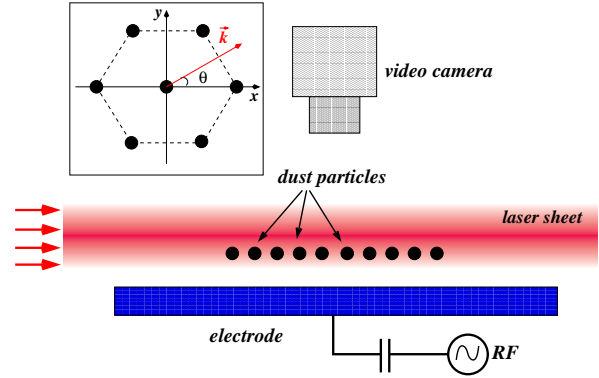}
\caption{(Color online) Sketch of the experimental setup. Microparticles are confined above the RF electrode and are illuminated
with a horizontal laser sheet having a Gaussian profile in the vertical direction. The monolayer is levitated well below the peak
of the laser intensity, which results in strong intensity variations of the scattered light upon the vertical displacement of
individual particles. The upward/downward displacement corresponds to positive/negative intensity variation. Inset shows
elementary cell of the hexagonal lattice and the frame of reference chosen in this paper, the orientation of the wave vector
${\bf k}$ is measured in respect to the $x$ axis.}
\label{fig:schematic}
\end{figure}

A sketch of the experimental setup is shown in Fig.~\ref{fig:schematic}. We used a capacitively coupled RF glow discharge at
13.56 MHz. The argon pressure $p$ was between 0.5 Pa and 1 Pa and the RF peak-to-peak voltage $V_{\rm pp}$ was between 175 V
and 310 V (which corresponds to a forward RF power $P$ between 5 W and 20 W). The plasma parameters in the bulk discharge
were deduced from Langmuir probe measurements, yielding the electron temperature $T_e=2.5$~eV and the electron density
$n_e=2 \times10^9$~cm$^{-3}$ at $p=$0.66~Pa and $P$=20~W \cite{nosenko:2009}. A 2D particle suspension was formed by
levitating melamine formaldehyde microspheres in the sheath above the RF electrode. The particles had a diameter of $8.77\pm0.14~\mu$m and
a mass $m=5.3\times10^{-13}$~kg. The diameter of the obtained crystalline structure was about 50--60~mm, depending on the
number of injected particles. The microparticles were illuminated by a horizontal laser sheet which had a Gaussian profile
in the vertical direction with a standard deviation $\sigma\simeq 75~\mu$m (corresponding to a full width at half maximum of 175
$\mu$m). The sheet thickness was approximatively constant across the whole  crystal. The particles were imaged through a window at the top of the chamber by a Photron FASTCAM 1024 PCI camera at a
speed of 250 frames per second. The horizontal coordinates $x$ and $y$ as well as velocity components $v_x$ and $v_y$ of
individual particles were then extracted with sub-pixel resolution in each frame by using a standard particle tracking
technique \cite{1478-3975-4-3-008}. An additional side-view camera was used to verify that our experiments were carried out
with a single layer of particles.

In order to extract the vertical position $z$ and velocity $v_z$ of individual particles, we needed to employ a very different
technique. Unlike relatively compact Coulomb clusters, where these values can be directly measured by using the side view,
the particles forming large monolayers cannot be individually resolved from the side. Therefore, the out-of-plane particle
tracking can only be performed by using the top view, and then the vertical displacement should be deduced from the relative
variation of the scattered light intensity.

The ``conventional'' method of particle visualization, when the peak of the Gaussian intensity in the illuminating laser
sheet practically coincides with the particle monolayer (this facilitates the in-plane tracking, especially at higher frame
rates), is not appropriate to obtain $z$ and $v_z$. The reason is twofold: The first problem of such configuration is that
the variation in the (particle image) intensity between consecutive frames gives us only the magnitude of the vertical
displacement (and hence velocity), whereas the direction remains undefined. Indeed, the intensity of the individual particle
image at each frame is proportional to the local illumination intensity, which scales as
\begin{equation}\label{eq:eq1}
I(z) \propto e^{-(z-z_{\rm max})^2/2\sigma^2},
\end{equation}
where  $z_{\rm max}$ is the position of the intensity maximum. Therefore, if the mean levitation height $z_{\rm lev}$ is very
close to $z_{\rm max}$, then the magnitude of the particle displacement $|\delta z|$ can be substantially larger than
$|z_{\rm lev}-z_{\rm max}|$. Obviously, a given variation of the image intensity in this case may equally imply either
positive or negative displacement \cite{samsonov:026410}. The second problem of the ``conventional'' method is associated
with the sensitivity. Since $\delta z\gtrsim|z_{\rm lev}-z_{\rm max}|$, the resulting intensity variations scale as $\delta
I/I\approx\frac12(\delta z/\sigma)^2$. The magnitude of vertical displacement (induced by thermal fluctuations) can be
estimated as $|\delta z|\sim\sqrt{T_d/m_d\Omega_{\rm v}^2}\lesssim10~\mu$m (here $T_d$ is the kinetic temperature of
particles and $\Omega_{\rm v}=2\pi f_{\rm v}$ is the resonance frequency in their vertical confinement). This value is much
smaller than the standard deviation $\sigma\approx 75~\mu$m. Thus, we conclude that the relative intensity variations are at
the level of $\sim 1~\%$, which makes them very difficult to detect.

\begin{figure}
\includegraphics[width=8.3cm]{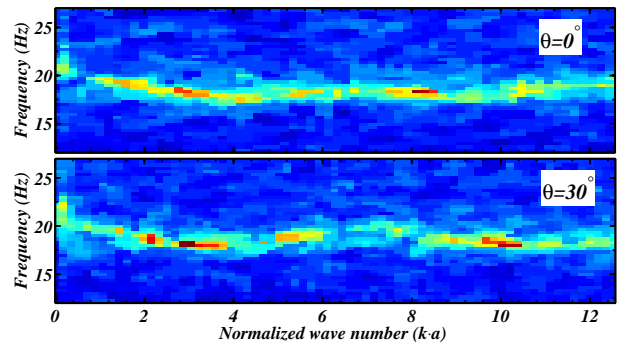}
\caption{(Color online) Fluctuation spectra of the naturally excited out-of-plane DL waves in a 2D crystal. The fluctuation
intensity is color-coded linearly, from dark-blue ($10^{-3}$ a.u.) to dark-red (1 a.u.), revealing the optical dispersion of the
mode. The results are for a pressure of 0.8~Pa and a RF power of 15~W. The upper and lower panels depict the
fluctuation spectra with the wave vector ${\bf k}$ pointed along the principal lattice axes, $\theta=0^{\circ}$ and
$\theta=30^{\circ}$, respectively (see Fig.~\ref{fig:schematic}). The wave vector is normalized by the inverse lattice
constant $a^{-1}$.}\label{fig:disp}
\end{figure}

In our experiments we employed an alternative method of particle visualization. The position of the intensity maximum was set
about $100~\mu$m above the levitation height, so that $|z_{\rm lev}-z_{\rm max}|/|\delta z|\approx 10$. This allowed us to
overcome both drawbacks mentioned above: (i) particles moved in the region where the vertical gradient of the illumination
intensity remained positive (which allowed us to avoid ambiguity in determining the direction of the displacement), and (ii)
the relative intensity variations scaled almost \textit{linearly} with the displacement, $\delta I/I\approx|z_{\rm
lev}-z_{\rm max}|\delta z/\sigma^2$, so that the resulting magnitude of $\delta I/I$ was about 15 times larger than that in
the ``conventional'' tracking configuration. It is worth noting that this alternative visualization method does not
critically affect the quality of the tracking in the horizontal $xy$ plane compared to the ``conventional'' tracking 
configuration. A separate test has indeed shown that the pixel locking effect is roughly at the same level and that 
the \textit{in-plane} fluctuation spectra are similar. 

By performing the Fourier transform in time and space of the determined particle velocities, the dispersion relations
$f({\bf k})$ (both in-plane and out-of-plane) can then be obtained for a 2D crystal.

The resulting fluctuation spectra of the out-of-plane waves are shown in Fig.~\ref{fig:disp} for two principal orientations
of the wave vector, at $\theta=0^{\circ}$ and $30^{\circ}$. The spectra represent the wave energy distribution in the
$(f,{\bf k})$ space, so that the ``ridge'' of this distribution yields the wave dispersion relation. The most conspicuous
feature of the measurements is the optical character of the dispersion relation -- the frequency of the long-wavelength
waves is finite. Moreover, the dispersion at long wavelengths is negative, i.e., the wave frequency falls off as the
wave number increases. At larger $|{\bf k}|$, the wave dispersion is different for $\theta=0^{\circ}$ and $30^{\circ}$.

Experimental observation of the out-of-plane wave dispersion is the main result of our paper. Figure~\ref{fig:disp} shows
that the wave frequency as the function of the wave vector ${\bf k}$ changes in a narrow interval of $17$~Hz~$\lesssim
f\lesssim20$~Hz, which makes resolving the wave dispersion a challenging task. Our method, however, allowed us to clearly
identify the wave dispersion. Indeed, the frequency resolution of $\approx 1.5$~Hz (defined as the standard deviation of a
Gaussian fit) is smaller than the measured wave frequency range $\Delta f(\approx3~{\rm Hz})$.

\begin{figure}
\includegraphics[width=8.3cm]{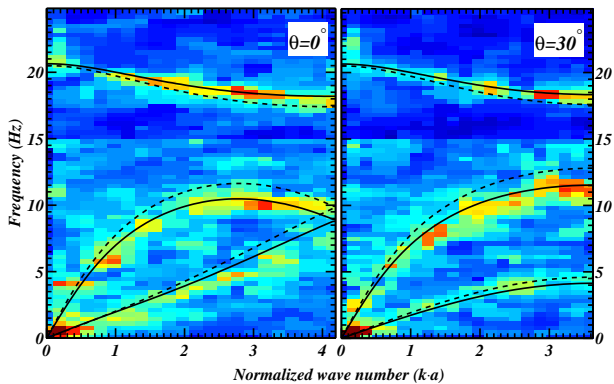}
\caption{(Color online) Comparison of the experimentally measured DL dispersion relations with theory. Shown are the 2D
in-plane and out-of-plane dispersion relations deduced from the experiments (at RF power of 15~W and pressure of 0.8~Pa) and
the theoretical curves.
The results are presented for two directions of the wave vector ${\bf k}$ (for $\theta=0^{\circ}$ and $\theta=30^{\circ}$, see
Fig.~\ref{fig:schematic}). The dashed lines are theoretical dispersion relations for pure Yukawa interaction between particles
(with the same parameters as in the experiment, see Table~\ref{tab:parameters1}), the solid lines represent the case when the
interaction with plasma wakes is taken into account. The shown range of $|{\bf k}|$ is limited by the first
Brillouin zone and therefore depends on the direction: $|{\bf k}|a=\frac43\pi$ for $\theta=0^{\circ}$ and $|{\bf
k}|a=\frac2{\sqrt{3}}\pi$ for $\theta=30^{\circ}$. The wave vector is normalized by the inverse lattice constant $a^{-1}$.}
\label{fig:allmodep15}
\end{figure}

Our method allowed us to measure all three wave modes of a 2D plasma crystal in a single experimental run. In
Fig.~\ref{fig:allmodep15} we show the fluctuation spectra of out-of-plane and in-plane wave modes obtained from the same
experiment as in Fig.~\ref{fig:disp}, for two principal lattice orientations, $\theta=0^{\circ}$ and $30^{\circ}$. Only the
first Brillouin zone is shown here. Lines represent theoretical dispersion relations which depend on two lattice parameters
-- particle charge number $Z_d$ and the lattice screening parameter $\kappa=a/\lambda$, where $a$ is the interparticle
distance (lattice constant) and $\lambda$ the effective screening length. The theoretical dispersion relations were
calculated using the following procedure. First, the lattice parameters were deduced from the low-$|{\bf k}|$ part of
in-plane fluctuation spectra using the method of Ref.~\cite{PhysRevLett.89.035001}. The obtained values of $Z_d$ and
$\kappa$ are shown in Table~\ref{tab:parameters1} for four different experiments. Second, these values were used in the
theory of Ref.~\cite{zhdanov:zim} to calculate the whole dispersion curves (the difference between the dashed and solid curves
will be discussed later).

\begin{table}
\caption{Parameters of a 2D crystal extracted from different experiments (marked from I to IV). The lattice constant $a$ is
obtained from the position of the first peak of the pair correlation function. Lattice parameters $Z_d$ and
$\kappa=a/\lambda$ are deduced from the in-plane phonon spectra with accuracy $\approx15~\%$ and $\approx30~\%$,
respectively. The accuracy for the measured vertical resonance frequency $f_{\rm v}$ is $\approx8~\%$. The frequency range
$\Delta f$ is the difference between $f_{\rm v}$ and the minimum frequency of the out-of-plane mode at $\theta=0^{\circ}$
(see Fig.~\ref{fig:schematic}). Subscripts ``exp'' and ``th'' stand for ``experimental'' and ``theoretical'', respectively,
${\Delta f}_{\rm th}$ is calculated using the theory of Ref.~\cite{zhdanov:zim}.} \label{tab:parameters1}
\begin{center}
\begin{tabular}{c c c c c c c c c}
\hline \hline
 & $P$ & $p$ & $a$ & $|Z_d|$ & $\kappa$ & $f_{\rm v}$ & $\Delta f_{\rm exp}$ & $\Delta f_{\rm th}$\\
 & (W) & (Pa) & ($\mu$m) & ({\it e}) &  & (Hz) & (Hz) & (Hz) \\
\hline
I & 20    &0.8 & 525 & 14000 & 1.3 & 23.3 &2.7 & 1.9\\
II & 15    &0.8 & 514 & 14200 & 1.0 & 20.5 &2.9 & 3.1\\
III & 5     &0.8 & 1060 & 14600 & 1.3  & 18.3 &0.6 & 0.4 \\
IV & 5     &0.9 & 600 & 18000 & 1.0 & 16.3 & 3.7 & 3.9\\
\hline\hline
\end{tabular}
\end{center}
\end{table}

As can be seen in Fig.~\ref{fig:allmodep15}, all three branches of the experimental wave spectra can be simultaneously
fitted by theoretical curves calculated with the same set of parameters. This provides a useful consistency check for our
experimental method and also for theory. Good overall agreement of our measurements and the theory of
Ref.~\cite{zhdanov:zim} serves as a verification that our method correctly measures the out-of-plane motion of particles and
also as an indication that the theory of Ref.~\cite{zhdanov:zim} correctly captures the essential physics of 2D plasma
crystals.

The out-of-plane wave mode can also be useful for diagnostic purpose. The lattice parameters $Z_d$ and $\kappa$ are usually
determined by fitting in-plane dispersion relations \cite{PhysRevLett.89.035001}. Another approach is to measure the
propagation speed of externally excited pulses \cite{PhysRevLett.88.215002}. But under certain conditions (e.g., near
the melting line) it is not practical and one must rely on natural waves and,  in this case, in-plane fluctuation spectra  can be very noisy. Therefore, the
additional fitting of the out-of-plane dispersion relation provides an important complementary tool to check consistency of
the measurements. As an example, in Table~\ref{tab:parameters1} we compare the experimental frequency range for the
out-of-plane mode $\Delta f_{\rm exp}$ to the theoretically calculated one $\Delta f_{\rm th}$ \cite{zhdanov:zim}. One can
see that experimental and theoretical values agree well in most cases (except of experiment III that we discuss later),
which is a strong indication that the crystal parameters were measured properly.

Next, we discuss how the out-of-plane waves are affected by the size of the plasma crystal. Theory
\cite{PhysRevE.63.016409,PhysRevE.56.R74,zhdanov:zim,PhysRevE.68.026405} suggests that the frequency of the out-of-plane
mode varies between $f_{\rm v}$ and $f_{\rm v}-\Delta f$, in a narrow frequency range $\Delta f$. It scales with the lattice
constant as $\Delta f_{\rm th}\propto \Omega_{\rm DL}^2 \propto a^{-3}$, where $\Omega_{\rm DL}^2=e^2Z_d^2/m_da^3$ is the
(squared) frequency scale of DL waves (it is assumed that $2\pi f_{\rm v}\gg\Omega_{\rm DL}$). The exact factor that
determines the magnitude of $\Delta f_{\rm th}$ has a relatively weak dependence on $\kappa$ (which, in turn, only varies
within 30~\% in our experiments) and can be obtained from the rigorous theory for a 2D crystal ~\cite{zhdanov:zim}. Thus,
for bigger crystals where the lattice constant $a$ is smaller (experiments I, II, and IV in Table~\ref{tab:parameters1}),
the phonon frequency range $\Delta f$ is larger, which makes it easier to resolve in an experiment. In addition, in big
crystals the central part that is used for data analysis is fairly homogeneous. In contrast, for small crystals (experiment
III in Table~\ref{tab:parameters1}) the situation is the opposite. The frequency range $\Delta f \propto a^{-3}$ is small
and the fluctuation spectrum is broadened due to the significant variation of $a$ within the analysis area. This clearly
results in greater discrepancy between experiment III and theory, up to $50$~\%. Thus, the out-of-plane wave mode can be
better studied using bigger crystals.

It is worth mentioning that theory \cite{PhysRevE.63.016409,zhdanov:zim,PhysRevE.68.026405} predicts that both the
out-of-plane and in-plane modes are affected by the presence of plasma wakes. If we introduce the effective positive charge
$q$ accumulated in the wake (normalized by the absolute value of particle charge), then the frequency of the in-plane modes
is diminished by a factor $\sqrt{1-q}$, whereas the frequency range of the out-of-plane mode, $\Delta f$, is
proportional to $(1-q)$. Figure~\ref{fig:allmodep15} illustrates this effect: The dashed curves are calculated for a crystal
with a pure Yukawa interaction between particles, whereas the solid curves are for the case when the additional
particle-wake interactions are taken into account (with $q=0.2$) \cite{PhysRevE.63.016409,zhdanov:zim,PhysRevE.68.026405}. This ``wake
correction'' apparently improves agreement with the experiments.

Direct observations of the atomistic dynamics in ``regular'' 2D or quasi-2D systems (such as carbon nanotubes
\cite{popov:085407} and graphite \cite{PhysRevLett.88.027401}) are inhibited, and therefore experimental measurements of
phonon spectra in such systems rely on indirect methods, e.g., Raman scattering technique \cite{PhysRevB.59.62,
PhysRevLett.88.027401}. Natural model systems, such as crystalline complex plasmas, provide us with a unique opportunity to
study generic properties of wave modes sustained in classical strongly coupled ensembles, explore limits of small
``nano''-clusters, investigate peculiarities of the atomistic dynamics upon nonlinear mode coupling, etc. 
As compared to previously reported investigations of wave modes in 2D plasma crystals, the results presented in this Letter 
show a substantial improvement of the  particle imaging technique. It allowed us for the fist time to directly observe the optical phonon spectrum 
which has been predicted a long time ago. This development
gives us strong grounds to believe that the direct observation of optical modes will allow 
 much better discrimination of various subtle mechanisms affecting wave modes in 2D crystalline systems.

The authors thank U. Konopka, B. Klumov, and D. Samsonov for valuable discussions.

\end{document}